\documentclass[twocolumn,showpacs,preprintnumbers,amsmath,amssymb]{revtex4}
\usepackage{txfonts}
\usepackage{}
\usepackage{dcolumn}
\usepackage{bm}
\usepackage{graphicx}
\usepackage{amsmath}
\begin{document}

\title{Coherent control via interplay between driving field and two-body interaction in a double well}
\author{Juan Liu,\ Wenhua Hai\footnote{Corresponding author. Email
address: whhai2005@yahoo.com.cn},\ Zheng Zhou} \affiliation{Department of physics
and Key Laboratory of Low-dimensional Quantum Structures and
\\ Quantum Control of Ministry
of Education, Hunan Normal University, Changsha 410081, China}

\begin{abstract}
We investigate interplay between external field and interatomic
interaction and its applications to coherent control of quantum
tunneling for two repulsive bosons confined in a high-frequency
driven double well. A full solution of the system is generated analytically as a coherent non-Floquet state by using the Floquet
states as a set of complete bases. It
is demonstrated that the photon resonance of interaction leads
to translation of the Floquet level-crossing points, and the non-resonant
interaction causes avoided crossing of partial levels. In the
non-Floquet states, the bosons beyond the crossing points slowly
vary their populations, and the resonant (non-resonant) interactions
enhance (decrease) the tunneling rate of the paired particles. Three
different kinds of the coherent destructions of tunneling (CDT) at
the crossing, avoided-crossing and uncrossing points, and the
corresponding stationary-like states, are illustrated. The analytical
results are numerically confirmed and perfect agreements are found.
Based on the results, an useful scheme of quantum tunneling switch
between stationary-like states is presented.

\end{abstract}

\pacs{32.80.Qk, 03.65.Ge, 03.65.Xp, 05.30.Jp }

\maketitle

\section{Introduction}

Coherent control of quantum tunneling in a double well via
periodical driving has been researched extensively from both
theoretical and experimental sides \cite{Milena,Kohler,P,E.Kierig}.
Some interesting phenomena, such as the coherent destruction or
construction of tunneling (CDT or CCT)
\cite{E.Kierig,G.Della,F.Grossmann,Xie,Gong,Longhi1,Hai1}, chaos
enhancing tunneling \cite{Lin, LuG}, and photon-assisted tunneling
\cite{C.Sias,Q} have been found. Many works focus on single- or
many-particle systems. Few-particle systems are of a intermediate
class between the both systems, which deserve further study for
comprehensively understanding tunneling dynamics. However,
investigations on quantum control to two particles in a periodically
driven double well are extremely rare, except for the cases of two
particles in a one-dimensional lattice which can be reduced to a two-site trap \cite{C.E, Longhi2} or two
particles in a driven
double-well train \cite{Creffield,Romero,Hai,M.Esmann}. Recently,
some relevant researches have been completed for a non-driven
few-particle system \cite{D.S,P.Cheinet,Budhaditya}. Several
interesting phenomena of quantum tunneling were shown for two
repulsive bosons in a non-driven double well, which include the Rabi
oscillations and correlated pair tunneling
\cite{Sascha,Longhi,Sascha2}. The interatomic interaction adjusted
by the Feshbach- resonance technique \cite{M.Z} plays an important
role in tunneling dynamics of the non-driven two-particle system.
Here we are interested in the combined effects of the driving and
interaction on  tunneling dynamics of double-well coupled two
bosons.

In this paper, we study coherent control of quantum tunneling via
the competition and cooperation between atomic interaction and
driving field for a pair of repulsive bosons confined in a
periodically driven double well. In the high-frequency regime, we
obtain a set of Floquet quasienergies and Floquet states of
invariant population, which contain the quasi-NOON state, an
interesting entanglement state \cite{Chen,K.Stiebler}. The general
non-Floquet state of slowly varying population is generated as a full solution which is
a coherent superposition of the Floquet states. The quasienergies as
function of the driving parameters are plotted for several different
values of interaction strength. The quasienergy spectra exhibit that
comparing with the noninteracting case, the photon resonance
\cite{C.E} leads to translation of the level-crossing points, and
the non-resonant interactions cause avoided crossing of partial
levels. Subsequently, exploiting the coherent non-Floquet solutions,
we investigate time evolution of the particle population and
demonstrate that beyond the crossing points, the bosons slowly vary
their populations compared to the high-frequency driving. The
resonant or non-resonant interactions can enhance or decrease the
tunneling rate of the paired particles. Three different kinds of CDT, respectively
at the level-crossing points for the superposition of three states,
at the avoided-crossing points for the superposition of two states
and at the uncrossing points for the single Floquet states, are
illustrated, which result in different stationary-like states of
invariant populations. The analytical results are confirmed by
direct numerical simulations and good agreements are shown. As an
application of the above results, an interesting scheme of quantum
tunneling switch between stationary-like states is presented by
using CDT or CCT to close or open the quantum tunneling, which could
be useful for the quantum control of two bosons in a double well.

\section{Analytical solutions in high-frequency regime}

We consider two ultracold bosons confined in a periodically driven double well with the governing Hamiltonian \cite{Yosuke,D.Jaksch}
\begin{eqnarray}\label{dy1}
H(t)&=&\frac{\varepsilon(t)}{2}(b^\dag_{1}b_{1}-b^\dag_{2}b_{2})+\gamma(b^\dag_{1}b_{2}+b^\dag_{2}b_{1})\nonumber\\
&+&
\frac{U}{2}[b^\dag_{1}b_{1}(b^\dag_{1}b_{1}-1)+b^\dag_{2}b_{2}(b^\dag_{2}b_{2}-1)],\end{eqnarray}
where $b_{i} (b_{i}^{\dag})$ for $i=1,2$ are the annihilation
(creation) operators in the $i$th localized state which may be a line superposition of the lowest doublet of single-particle
energy eigenstates \cite{Jinasundera}. Their commutation relations
read $[b_{i},b_{j}^{\dag}]=\delta_{ij}$. The parameters $\gamma$ and
$U$ are the tunneling coefficient between two wells and the
interaction strength between two bosons. The function
$\varepsilon(t)=\varepsilon_{0}\cos(\omega t)$ describes the
external $ac$ field in which $\omega$ and $\varepsilon_{0}$ are the
driving frequency and amplitude, respectively. For simplicity, we
have put $\hbar=1$ and taken the reference frequency
$\omega_{0}=10^{2}s^{-1}$ \cite{MHolthaus} so that
$\varepsilon_{0}$, $\gamma$, $\omega$ and $U$ are in units of
$\omega_{0}$, and time $t$ is normalized in units of
$\omega_{0}^{-1}$.

Quantum state $|\psi(t)\rangle$ of system (1) can be
expanded in Fock bases $|0,2\rangle$,$|1,1\rangle$,$|2,0\rangle$ as \cite{C.E}
\begin{eqnarray}\label{dy2}
|\psi(t)\rangle&=&a_{0}(t)|0,2\rangle+a_{1}(t)|1,1\rangle+a_{2}(t)|2,0\rangle,
\end{eqnarray}
where $|j\rangle=|j,2-j\rangle (j=0,1,2)$ denote that $j$ bosons reside in
left well, $2-j$ bosons reside in right well; $a_{j}(t)$
represents probability amplitude of the system in $j$-th Fock state $|j\rangle$,
which obey the normalization condition
$|a_{0}(t)|^{2}+|a_{1}(t)|^{2}+|a_{2}(t)|^{2}=1$. Inserting Eqs. (1)
and (2) into the time-dependent Schr\"{o}dinger Equation
$i\frac{\partial|\psi(t)\rangle}{\partial t}=H(t)|\psi(t)\rangle$ results in the
coupling equations
\begin{eqnarray}\label{dy3}
i\dot{a}_{0}(t)=[U-\varepsilon(t)]a_{0}(t)+\sqrt{2}\gamma a_{1}(t),\nonumber\\
i\dot{a}_{1}(t)=\sqrt{2}\gamma a_{0}(t)+\sqrt{2}\gamma a_{2}(t),\nonumber\\
i\dot{a}_{2}(t)=[U+\varepsilon(t)]a_{2}(t)+\sqrt{2}\gamma a_{1}(t).
\end{eqnarray}
It is difficult to get exact analytical solutions of Eq. (3) with periodic driving $\varepsilon(t)=\varepsilon_{0}\cos(\omega
t)$. However, we can construct the approximate analytical solution in the high-frequency limit, $\omega\gg 1$ and $u, \gamma\ll\omega$. To do this, we adopt the idea of reduced interaction strength \cite{C.E} to rewrite the interaction strength as $U=n\omega+u$ for $0\leq u\ll\omega,\ n=0,1,2,...$ with $u$ being the reduced interaction strength, and make the function
transformation $a_{0}(t)=b_{0}(t)e^{i\int (\varepsilon(t)-n
\omega)dt}$, $a_{1}(t)=b_{1}(t)$,
$a_{2}(t)=b_{2}(t)e^{-i\int(\varepsilon(t)+n \omega) dt}$, where
$b_{j}(t)(j=0,1,2)$ are slowly-varying functions of time, then Eq. (3) becomes new equations in terms of $b_j(t)$. Because of $|b_j(t)|=|a_j(t)|$, the high-frequency limit implies that the occupied probabilities of Fock states slowly vary in time. Exploiting
Fourier expansion $\exp[\int(\pm
i\varepsilon(t)-in\omega)dt]=\exp[\pm
i(\frac{\varepsilon_{0}}{\omega})\sin(\omega t) -in \omega
t]=\sum_{n^{'}=-\infty}^{\infty}\mathcal{J}_{n^{'}}(\frac{\varepsilon_{0}}{\omega})\exp[
i(\pm n^{'}-n)\omega t]$, we easily obtain the time-average of the rapidly oscillating function as $\mathcal{J}_{\pm n}(\frac{\varepsilon_{0}}{\omega})$ which are the $\pm n$-order bessel
function. Under the
high-frequency approximation, the rapidly oscillating function can be
replaced by its time-averaging value \cite{Yosuke} such that the equations of $b_j(t)$ are
transformed into
\begin{eqnarray}\label{dy4}
i\dot{b}_{0}(t)=u
b_{0}(t)+J_{n}(\frac{\varepsilon_{0}}{\omega})b_{1}(t),\nonumber\\
i\dot{b}_{1}(t)=
J_{n}(\frac{\varepsilon_{0}}{\omega})b_{0}(t)+(-1)^{n}
J_{n}(\frac{\varepsilon_{0}}{\omega})b_{2}(t),\nonumber\\
i\dot{b}_{2}(t)=u
b_{2}(t)+(-1)^nJ_{n}(\frac{\varepsilon_{0}}{\omega})b_{1}(t).
\end{eqnarray}
In the calculations, we have employed the formula $\mathcal
{J}_{-n}(\frac{\varepsilon_{0}}{\omega})=(-1)^{n}\mathcal
{J}_{n}(\frac{\varepsilon_{0}}{\omega})$ for positive
integer $n$, and the renormalized coupling coefficient $J_{n}(\frac{\varepsilon_{0}}{\omega})=\sqrt{2}\gamma
\mathcal {J}_{n}(\frac{\varepsilon_{0}}{\omega})$. Starting from Eq. (4), we obtain the interesting analytical solutions as follows.

\subsection{Quasienergies and Floquet states}

Because the time-dependent Hamiltonian (1) has the period
$T=\frac{2\pi}{\omega}$, we can make use of Floquet
theory \cite{Sambe} to get its solution in the form
$|\psi(t)\rangle=e^{-iEt}|\varphi(t)\rangle$, where $|\varphi(t)\rangle$ is the Floquet state with the same period $\frac{2\pi}{\omega}$, $E$ is
called the Floquet quasienergy which has been normalized in units of
$\hbar \omega_{0}$. Noting the same period of the transformation function $e^{\int(\pm
i\varepsilon(t)-in\omega)dt}$ between $a_j(t)$ and $b_j(t)$, to generate the Floquet states, we seek the stationary solutions \cite{Lu} of Eq. (4)
$b_{0}=Ae^{-iEt}$, $b_{1}=Be^{-iEt}$, $b_{2}=Ce^{-iEt}$ with constants
$A,B,C$ obeying the normalization condition
$|A|^{2}+|B|^{2}+|C|^{2}=1$. Inserting these into Eq. (4), we establish the equations of the stationary solutions as
\begin{eqnarray}\label{dy5}
(E-u)A-J_{n}(\frac{\varepsilon_{0}}{\omega})B=0,\nonumber\\
-J_{n}(\frac{\varepsilon_{0}}{\omega})A+EB-(-1)^{n}J_{n}(\frac{\varepsilon_{0}}{\omega})C=0,\nonumber\\
-(-1)^{n}J_{n}(\frac{\varepsilon_{0}}{\omega})B+(E-u)C=0.
\end{eqnarray}
By solving Eq. (5), we obtain three Floqeut quasienergies $E_{l}$
and three sets of constants $A_{l}, B_{l}, C_{l}$ for $l=0,1,2$.
\begin{eqnarray}\label{dy6}
E_{0}=u, A_{0}=\frac{1}{\sqrt{2}},B_{0}=0,
C_{0}=(-1)^{n+1}\frac{1}{\sqrt{2}};
\end{eqnarray}
\begin{eqnarray}\label{dy7}
E_{1}&=&\frac{1}{2}(u-k_{n}), A_{1}=\frac{2J_{n}}{\sqrt{8J_{n}^{2}+(u+k_{n})^{2}}},\nonumber\\
B_{1}&=&-\frac{u+k_{n}}{\sqrt{8J_{n}^{2}+(u+k_{n})^{2}}}, C_{1}=(-1)^{n}\frac{2J_{n}}{\sqrt{8J_{n}^{2}+(u+k_{n})^{2}}};\nonumber\\
\end{eqnarray}
\begin{eqnarray}\label{dy8}
E_{2}&=&\frac{1}{2}(u+k_{n}), A_{2}=\frac{2J_{n}}{\sqrt{8J_{n}^{2}+(k_{n}-u)^{2}}},\nonumber\\
B_{2}&=&\frac{k_{n}-u}{\sqrt{8J_{n}^{2}+(k_{n}-u)^{2}}}, C_{2}=(-1)^{n}\frac{2J_{n}}{\sqrt{8J_{n}^{2}+(k_{n}-u)^{2}}}.\nonumber\\
\end{eqnarray}
Here, the simplified parameter
$k_{n}=\sqrt{8J_{n}^{2}(\frac{\varepsilon_{0}}{\omega})+u^{2}}$ has
been adopted, so that the Floquet quasiegergies are functions of
driving parameters and interaction strength. Obviously, for $u,
\gamma\ll\omega$ we have $E_{l}\ll\omega$. Thus as the functions of
$\exp[-iE_{l}t]$ the stationary solutions $b_j(t)$ are slowly
varying indeed. Given $b_j(t)$, the rapidly oscillating function
$a_{j}(t)$ are obtained immediately. Substituting such $a_{j}(t)$
into Eq. (2), Floquet states $|\psi_{l}(t)\rangle$ are expressed as
\begin{eqnarray}\label{dy9}
|\psi_{l}(t)\rangle&=&
e^{-iE_{l}t}(A_{l}e^{\frac{i\varepsilon_{0}}{\omega}\sin(\omega
t)-in\omega
t}|0,2\rangle+B_{l}|1,1\rangle\nonumber\\
&+&C_{l}e^{\frac{-i\varepsilon_{0}}{\omega}\sin(\omega t)-in\omega
t}|2,0\rangle)
\end{eqnarray}
for $l=0,1,2$. Here $n$ is a positive integer adjusted by the
interaction strength. If the system is prepared in the Floquet
states initially, probabilities of the system in different Fock
states are the constants $|A_{l}|^{2}$, $|B_{l}|^{2}$ and
$|C_{l}|^{2}$, respectively, so that tunneling of atoms between two
wells is suppressed completely, i.e., CDT occurs.

\subsection{Coherent non-Flouqet states}

Given the quasienergies and Floquet solutions of Eqs. (6-9), the
principle of superposition of quantum mechanics indicates that the
linear Schr\"odinger equation has the periodic or quasiperiodic
non-Floquet solutions, which is coherent superposition of the
Floquet states as a set of complete bases in a three-dimensional
Hilbert space \cite{Jinasundera, Longhi}. Noticing the relation
between the rapidly oscillating probability amplitude $a_j(t)$ and
the slowly varying function $b_j(t)$, the general non-Floquet state
has the form
\begin{eqnarray}\label{dy10}
|\psi(t)\rangle&=&c_{0}|\psi_{0}(t)\rangle+c_{1}|\psi_{1}(t)\rangle+c_{2}|\psi_{2}(t)\rangle\nonumber\\
&=&a'_{0}(t)|0,2\rangle+a'_{1}(t)|1,1\rangle+a'_{2}(t)|2,0\rangle\nonumber\\
&=&b'_{0}(t)e^{\frac{i\varepsilon_{0}}{\omega}\sin(\omega t)-in\omega
t}|0,2\rangle+b'_{1}(t)|1,1\rangle\nonumber\\
&+&b'_{2}(t)e^{-\frac{i\varepsilon_{0}}{\omega}\sin(\omega
t)-in\omega t}|2,0\rangle,
\end{eqnarray}
where $c_l$ $(l=0,1,2)$ are superposition coefficients determined by the initial conditions and normalization, and $b'_j(t)$ are the slowly varying functions
\begin{eqnarray}\label{dy11}
b'_{0}(t)=c_{0}A_{0}e^{-iE_{0}t}+c_{1}A_{1}e^{-iE_{1}t}+c_{2}A_{2}e^{-iE_{2}t},
\end{eqnarray}
\begin{eqnarray}\label{dy12}
b'_{1}(t)=c_{0}B_{0}e^{-iE_{0}t}+c_{1}B_{1}e^{-iE_{1}t}+c_{2}B_{2}e^{-iE_{2}t},
\end{eqnarray}
\begin{eqnarray}\label{dy13}
b'_{2}(t)=c_{0}C_{0}e^{-iE_{0}t}+c_{1}C_{1}e^{-iE_{1}t}+c_{2}C_{2}e^{-iE_{2}t}
\end{eqnarray}
with constants $A_{l}$, $B_{l}$, $C_{l}$ being given in Eqs.
(6)-(8). Clearly, the functions $a'_j(t)$ and $b'_j(t)$ obey Eqs. (3) and (4), respectively, and have the same norm which is the corresponding occupied probability. Any $b'_{j}(t)$ is a superposition of three periodic
functions with frequencies $\omega'_{l}=E_{l}$ for $l=0,1,2$. When all the frequency ratios
$\omega'_{l}/\omega'_{m}(l\neq m)$ for any pair of $l,m=0,1,2$ are rational numbers, any $b'_{j}(t)$
is a periodic function, and any irrational frequency ratio implies that all the
$b'_{j}(t)$ for $j=0,1,2$ are the quasi-periodic functions \cite{Lu}.

The non-Floquet state of Eq. (10) with Eqs. (11-13) is a general solution with constants $c_l$ being determined by the initial conditions and normalization. As a example, we consider the two atoms reside in right well initially, which means the normalized initial state of the system to be $|\psi(0)\rangle=|0,2\rangle$.
Inserting this into Eq. (10) yields
$b'_{0}(0)=1,b'_{1}(0)=b'_{2}(0)=0$, then substituting these and Eqs. (6-8) into
Eqs. (11-13) produces
\begin{eqnarray}\label{dy14}
b'_{0}(t)=\frac{1}{2}e^{-iut}+\frac{k_{n}-u}{4k_{n}}e^{-\frac{i}{2}(u-k_{n})t}
+\frac{k_{n}+u}{4k_{n}}e^{-\frac{i}{2}(u+k_{n})t},
\end{eqnarray}
\begin{eqnarray}\label{dy15}
b'_{1}(t)=-\frac{J_{n}}{k_{n}}e^{-\frac{i}{2}(u-k_{n})t}+\frac{J_{n}}{k_{n}}e^{-\frac{i}{2}(u+k_{n})t},
\end{eqnarray}
\begin{eqnarray}\label{dy16}
b'_{2}(t)&=&(-1)^{n+1}\frac{1}{2}e^{-iut}+(-1)^{n}\frac{k_{n}-u}{4k_{n}}e^{-\frac{i}{2}(u-k_{n})t}\nonumber\\
&+&(-1)^{n}\frac{k_{n}+u}{4k_{n}}e^{-\frac{i}{2}(u+k_{n})t}.
\end{eqnarray}
Combining Eq. (10) with Eqs. (14-16), we arrive at the special non-Floquet state associated with the initial state $|\psi(0)\rangle=|0,2\rangle$. Such a special state will be used, as an instance, to show the coherent control of quantum tunneling.

\section{Effects of driving and interaction on quasienergy}

We have already obtained the Floquet quasienergies, Floquet states
and non-Floquet states in high frequency regime. The general
non-Floquet state of Eq. (10) is a coherent superposition of two or
three Floquet states with two or three nonzero superposition
constants $c_l$ determined by the initial setup. It is worth noting
that in Eqs. (11-13) the three quasienergies appear in the
time-dependent phases of the probability amplitudes and directly
affect the tunneling probabilities through the phase coherence. When
the quasienergies of the Floquet states are different each other,
the tunneling probabilities are periodic or quasiperiodic functions
of time, which describe coherent population oscillation of the
system. If all the three quasienergies are the same, Eqs. (11-13)
mean invariant populations with constant $|b'_j(t)|$, namely the
level-crossing leads to CDT and stationary-like states. Similarly
for the case $E_0=E_2\ne E_1$, the superposition state of two
Floquet states $|\psi_0\rangle$ and $|\psi_2\rangle$ describes the
second kind of CDT, where the energy $E_1$ avoids the
level-crossing. Therefore, the analytical results render the the
relation between level-crossing and CDT more transparent and the
dependence of tunneling dynamics on the values of Floquet
quasienergies more evident. From the analytical results of Eqs.
(6-8) we know that the quasienergies are the functions of
interaction strength and driving parameters. In this section, we
study interplay between the external driving and interatomic
interaction, through the quasienergy spectra.

For a high frequency the expression $U=n\omega+u$ with $u\ll \omega$
implies that $n=0$ denotes noninteracting or weakly interacting
case, $n\ge 1$ correspond to strongly interacting case. Here, we
will consider only the cases $n=0$ and $n=1$, due to the interaction
$U=\omega+u$ strong enough. From Eqs. (6-8) the Floquet
quasienergies as functions of the driving parameters for several
different values of interacting strength $U$ and the same coupling
coefficients $\gamma$ are shown as the circles in Fig. 1.  We also
calculate the quasienergy spectra numerically from Eq. (3) for the
same parameters as those of the analytical results. The numerical
results are plotted as the solid curves of Fig. 1 and both the
results consistently illustrate the following interesting
phenomena.

\emph{Photon resonance leads to translation of the
level-crossing points}. In  Fig. 1(a) and Fig. 1(c) we show that for
the noninteracting and resonant interaction cases with $u=0$ and
$n=0,1$, quasienergy spectra exhibit exact level-crossings. Three
quasienergies are same at the crossing points, $E_0=E_1=E_2$, such
that the probabilities $|b'_{j}(t)|^{2}$ ($j=0,1,2$) of Eqs.
(11)-(13) do not change in time, and the CDT of the first kind
occurs. In the absence of interaction ($U=0$), the inset of Fig.
1(a) shows the first level-crossing location
$x_1=\varepsilon_{0}/\omega=2.405$ which is the first root of
equation $\mathcal{J}_0(x)=0$. While for the strong interaction
($U=50$) the inset of Fig. 1(c) indicates the first level-crossing
location $y_1=\varepsilon_{0}/\omega=3.832$ which is the first root
of $\mathcal{J}_1(y)=0$. The results mean that photon resonance
with $U=\omega$ results in translation of the location of
level-crossing point from $(\varepsilon_{0}/\omega,\ U)=(x_1,\ 0)$
to $(y_1,\ \omega)$. This implies that in the strong interaction case the onset
of CDT requires a greater $\varepsilon_{0}/\omega$ value, because of
$y_1 > x_1$.

\emph{Non-resonant interaction causes avoided crossing of partial
levels}. In Fig. 1(b) and Fig. 1(d) we show that for $u\neq0$ and $n=0,1$
the crossing of three levels is replaced by that of
two levels at the locations $(\varepsilon_{0}/\omega,\ U)=(2.405,\ 2)$ and
$(3.832,\ 52)$, respectively, namely the avoided crossing of one level appears. Such avoided
level-crossing points of partial levels correspond to driving
parameters obeying $\mathcal{J}_n(\varepsilon_{0}/\omega)=0$, which are the
points of closet approach of energies in quasi-energy spectra. At
such points the coherence between phases $E_0 t$ and $E_1 t$ may
exist in the non-Floquet state of Eq. (10) for the nonzero $c_l,\
l=0,1,2$. However, for the constants $c_1=0$ and $c_0,\ c_2\ne 0$,
the non-Floquet state becomes the superposition of two Floquet
states with the same quasienergy and time-dependent phase, so the
probabilities $|b'_{j}(t)|^{2} (j=0,1,2)$ of Eqs. (11-13) don't vary
in time and the second kind CDT occurs.

\begin{figure}[htbp] \centering
\includegraphics[height=1.2in,width=1.6in]{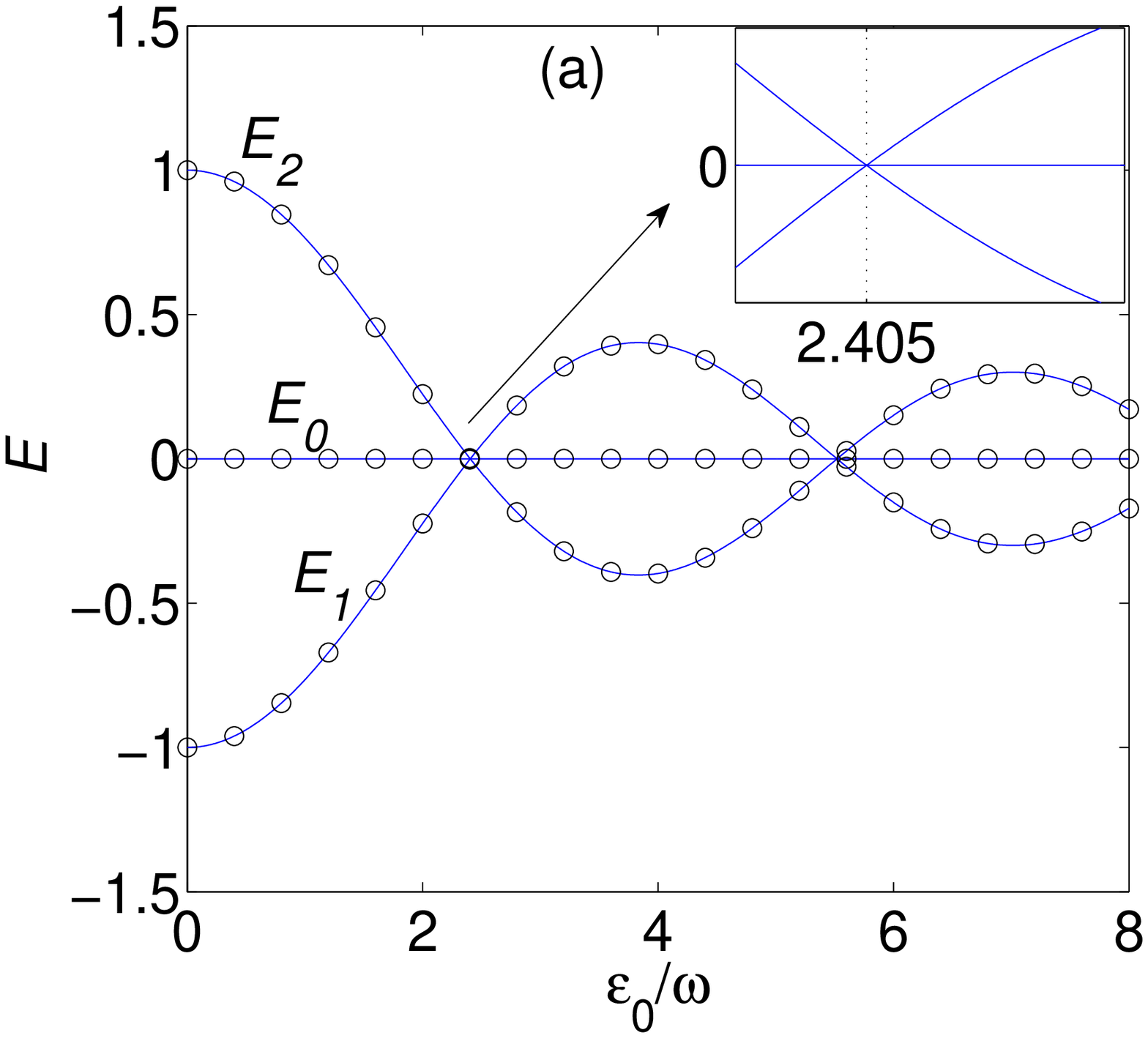}
\includegraphics[height=1.2in,width=1.6in]{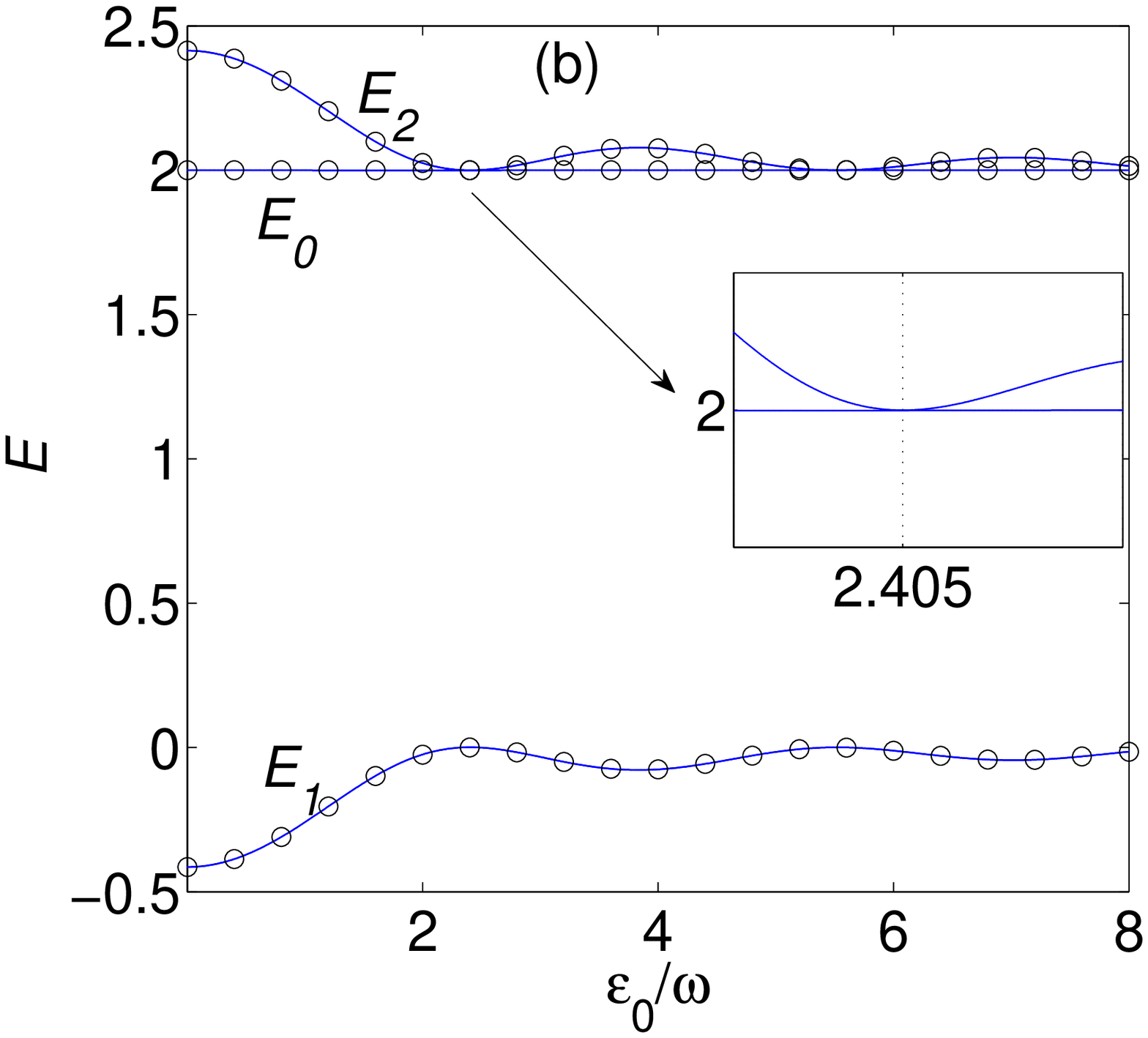}
\includegraphics[height=1.2in,width=1.6in]{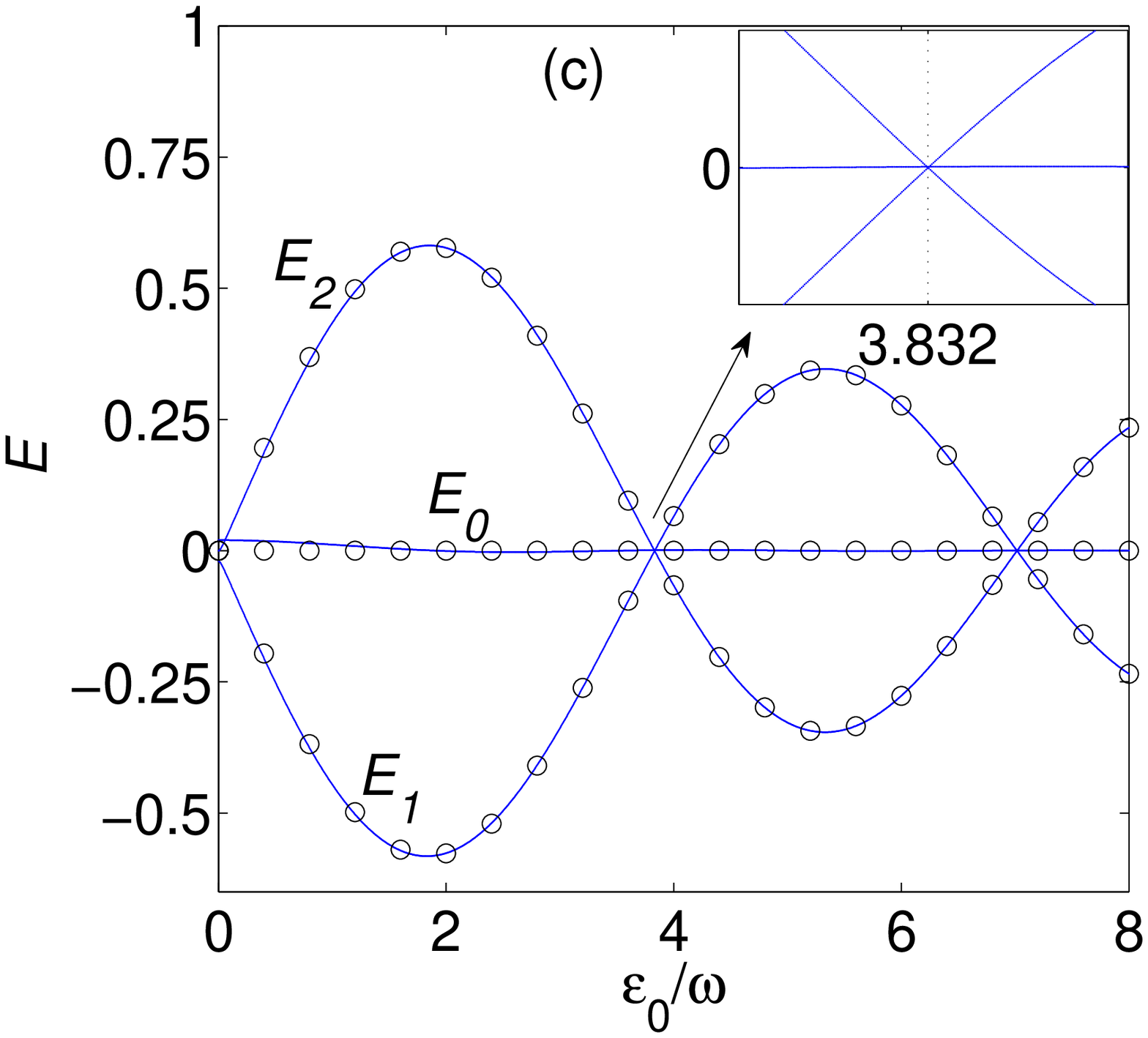}
\includegraphics[height=1.2in,width=1.6in]{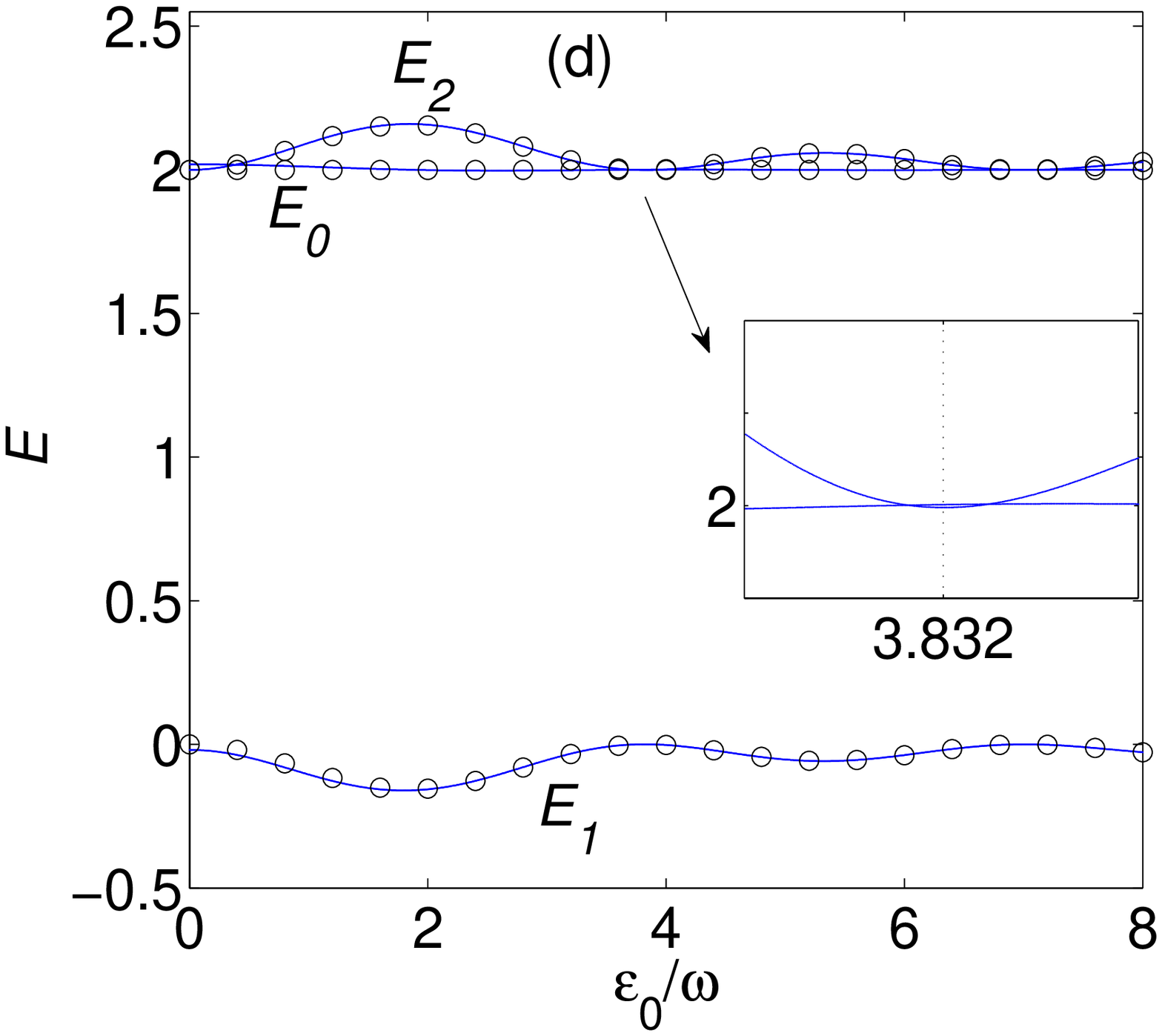}
\caption{\scriptsize{(Color online) Quasienergy spectra for the driving frequency $\omega=50$, coupling coefficient $\gamma=0.5$ and interaction strength (a) $U=0$; (b) $U=u=2$; (c) $U=\omega=50$; (d) $U=\omega+u=52$. Hereafter any quantity plotted in the figures is dimensionless, circles or circular points denote the analytical results and solid curves the numerical correspondences respectively, unless it is specially indicated.}}
\end{figure}

\begin{figure*}[htp]\center
\includegraphics[height=1.5in,width=2.2in]{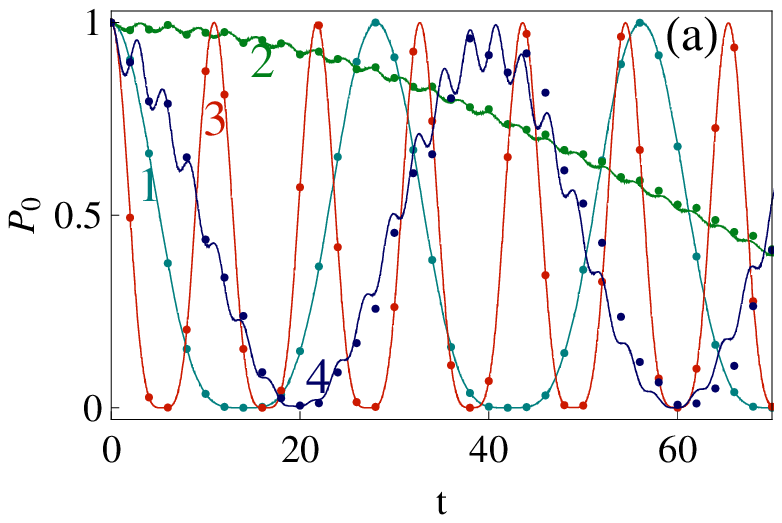}
\includegraphics[height=1.5in,width=2.2in]{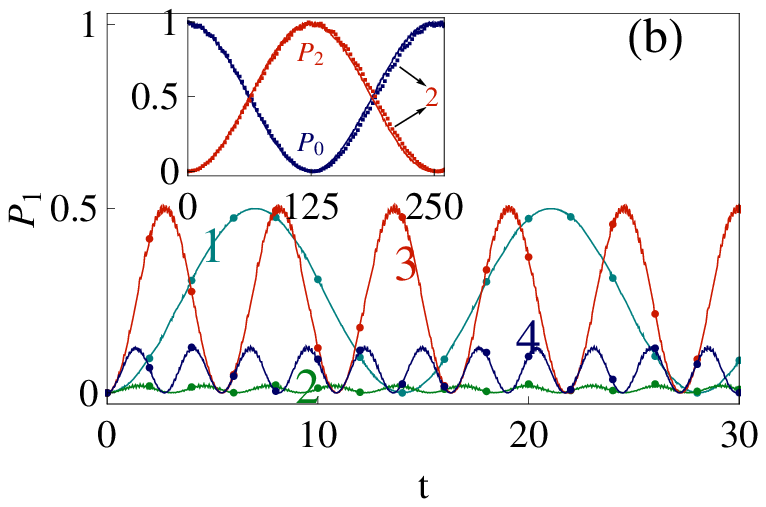}
\includegraphics[height=1.5in,width=2.2in]{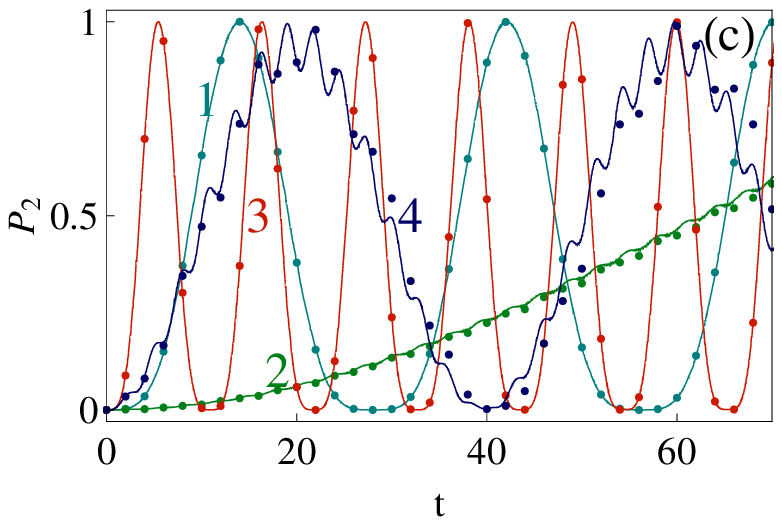}
\caption{\scriptsize{(Color online) Time evolutions of (a)
$P_{0}(t)$, (b) $P_{1}(t)$ and (c) $P_{2}(t)$
for the parameters $\varepsilon_{0}=100$, $\omega=50$ and $\gamma=0.5$, and $U=0$ (curve 1), $U=2$ (curve 2),
$U=50$, (curve 3) and $U=52$ (curve 4). The long-time
evolutions of curve $2$ in Figs. 2(a) and 2(c) are plotted in the inset of Fig. 2(b).}}
\end{figure*}

\section{Control and switch of quantum tunneling}

We have known that at the points without level-crossing the bosonic population oscillates periodically or quasi-periodically, and at the level-crossing points the initial population can be kept. Now we investigate the coherent control of quantum tunneling by applying the interplay between the external field and interatomic
interaction. For the level-uncrossing case we study how to coherently manipulate the tunneling rates by setting and adjusting the values of the syatem parameters. For the level-crossing and avoided crossing cases we perform control to the stationary-like states via CDT of different kinds. Then we apply these results to present a scheme for designing the quantum tunneling switch from a given state to different stationary-like states under CDT. Hereafter, all the analytical results are based on the general non-Floquet state of Eq. (10), and the initial conditions $b'_{0}(0)=1,b'_{1}(0)=b'_{2}(0)=0$ correspond to Eqs. (14-16) and other initial setups are associated with Eqs. (11-13).

\subsection{Manipulating tunneling rates}

At first, we consider the case in which the system parameters beyond
level-crossing points and the bosonic population oscillates periodically. We study how to control the tunneling rates by setting and adjusting the values of interaction strength and
driving parameters.

Letting $P_{j}(t)=|b'_{j}(t)|^{2}$ be probability of the system in
the $j$-th state, from Eqs. (14-16) we plot time evolutions of the
probabilities for $j=0,1,2$, $\omega=50$,
$\varepsilon_{0}=2\omega=100$ and $\gamma=0.5$, as shown in Fig. 2.
In this figure we show that all the probabilities are slowly varying
function of time comparing with the high-frequency driving. Defining
tunneling time as the needed time of the paired bosons tunneling
from state $|0,2\rangle$ to $|2,0\rangle$, a rough estimate from
curves 1, 2, 3 and 4 in Fig. 2(a) and the inset of Fig. 2(b)
indicates non-resonant tunneling time of the paired particles for
the weak interaction $U=2$ to be about $125$ and for the strong
interaction $U=52$ to be about $20$, compared to $6$ with resonant
interaction $U=50$ and $14$ without interaction. The results mean
that photon resonance ($U=\omega$) induces coherent
construction of tunneling (CCT) which leads to increase of the
tunneling rate of paired particles comparing with the non-resonance
case. Especially, for the weak interaction ($U=2$),
$P_{1}$ oscillates around $0$ nearly and its values are negligible. Thus the two bosons tunnel as pair between wells such that at any time, quantum
state of the system is a superposition of states $|0,2\rangle$ and
$|2,0\rangle$, which arrives at the NOON state with probabilities
$P_0=P_2=1/2$ periodically.

We also obtain numerical results from Eq. (3) with the same
parameters and initial conditions $a_{0}(0)=b'_{0}(0)=1,
a_{1}(0)=a_{2}(0)=b'_{1}(0)=b'_{2}(0)=0$ as those of the analytical
case, which are shown as solid lines of Fig. 2. The analytical
results are in good agreement with the numerical results except the
slight deviation for the case $U=52$.

\subsection{Preparing stationary-like states via CDT of different kinds}

When the tunneling rate is controlled to zero, CDT occurs and the
stationary-like state of invariant population is prepared. Such
stationary-like state may be a single Floquet state or the
superposition state of three or two Floquet states, which correspond
to CDT of three different kinds, respectively.

\emph{CDT at the level-crossing points}. At level-crossing point $(\varepsilon_{0}/\omega,\ U)=(2.405,\ 0)$ and for the high-frequency $\omega=50$, time evolution of $P_{0}(t)=|b'_{0}(t)|^{2}$ of Eq. (14)
is shown as curve 1 in Fig. 3(a). It is observed from this curve that $P_{0}(t)$ maintains the initial
value $P_{0}(0)=1$ so that tunneling between wells is suppressed completely and CDT of the first kind occurs.
Under the same driving parameters, for the level-uncrossing points with strong interaction $U=50$ and $52$, $P_{0}(t)$ still oscillates between
$0$ and $1$, as indicated by the curve 2 and curve 3 of Fig. 3(a). Then we consider the resonant case and tune the driving parameters to the level-crossing point $(\varepsilon_{0}/\omega,\ U)=(3.832,\ 50)$. The corresponding time evolution is shown by curve 4 in Fig. 3(b), where $P_{0}(t)$ is always equal to the initial value $1$,
indicating the occurrence of CDT. But at the level-uncrossing points
for non-interacting bosons with $U=0$ and weakly interacting
bosons with $U=2$, tunneling still exists, as displayed by the curve 5 and curve 6 of Fig. 3(b).

\emph{CDT at the avoided level-crossing points}. For the avoided level-crossing points of partial levels, $(\varepsilon_{0}/\omega,\ U)=(2.405,\ 2)$ and $(3.832,\ 52)$, from Eq. (14) the time evolutions of $P_{0}(t)$ are plotted as the coincided curve 1 and curve 4 of Fig. 3
respectively. The invariant probability $P_{0}(t)=1$ means CDT of the second kind, where Eq. (10) is the superposition state of $|\psi_0\rangle$ and $|\psi_2\rangle$ with the same energy $E_0=E_2$, due to $c_{1}=\lim\limits_{J_{n}\rightarrow
0}(k_{n}-u)\sqrt{8J_{n}^{2}+(k_{n}+u)^{2}}/(8J_{n}k_{n})=0$ for the given parameters.
The analytical results are confirmed numerically from Eq. (3)
for the same initial conditions and system parameters as those of the analytical calculations.
\begin{figure}[htbp]
\centering
\includegraphics[height=1.3in,width=2.2in]{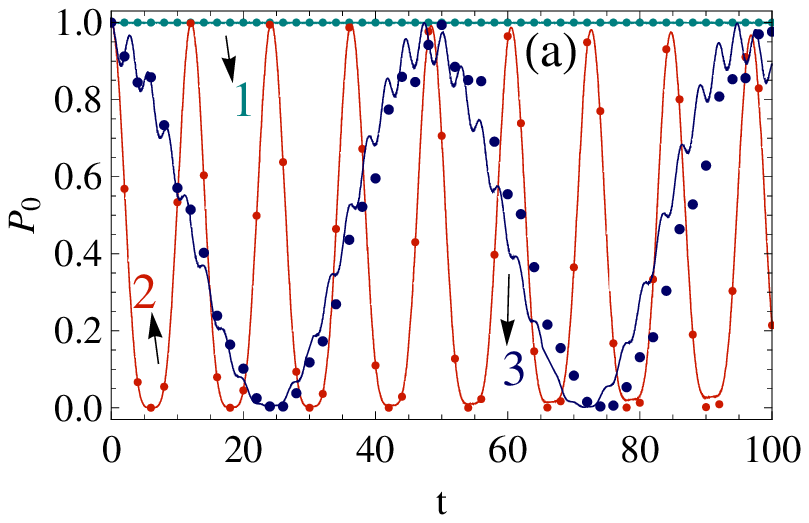}
\includegraphics[height=1.3in,width=2.2in]{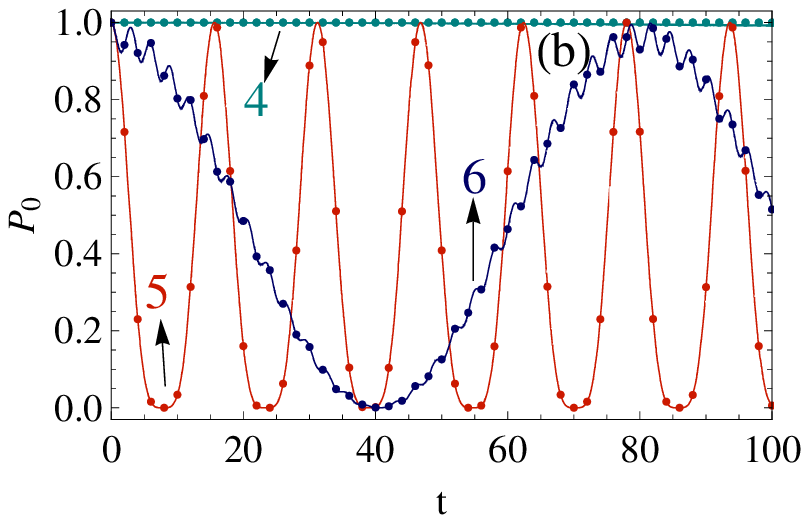}
\includegraphics[height=1.3in,width=2.2in]{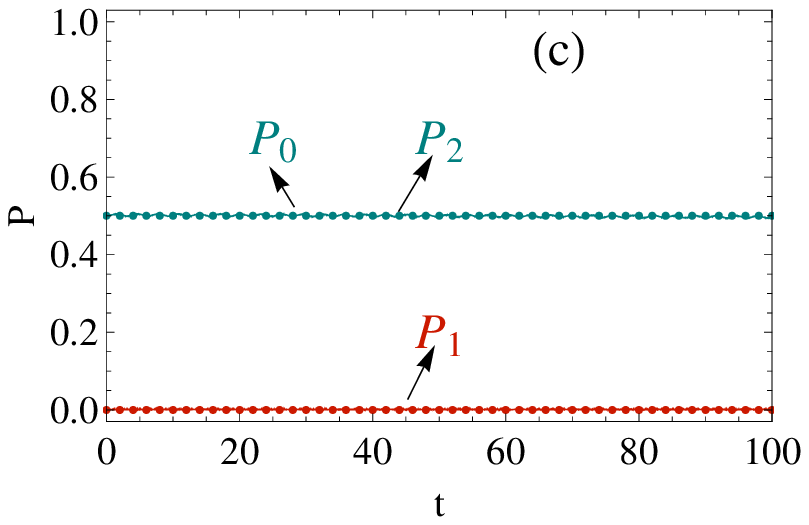}
\caption{\scriptsize{(Color online) Time evolutions of $P_{0}$ for the parameters
$\gamma = 0.5$, $\omega=50$, and (a)
$\frac{\varepsilon_{0}}{\omega}=2.405$, $U=0,\ 2$ (coincided curve 1),
$U=50$ (curve 2), $U=52$ (curve 3); (b) $\frac{\varepsilon_{0}}{\omega}=3.832$, $U=50,\ 52$ (coincided curve 4), $U=0$ (curve 5),
$U=2$ (curve 6); (c) $\frac{\varepsilon_{0}}{\omega}=2$, $U=2$ for the different initial
conditions $b'_{0}(0)=-b'_{2}(0)=1/\sqrt{2},\ b'_{1}(0)=0$.}}
\end{figure}

\emph{CDT for arbitrary values of the system parameters}. If the system is prepared
initially in a single Floquet state of Eq. (9), probability $P_{j}(t)$ of the
system in any $j$-th Fock state $|j,2-j\rangle$ is a constant for arbitrary values of the system parameters in high-frequency regime, including those at the points with level-crossing and without level-crossing. Such
an invariant population means CDT of the third kind. Based on such
CDT, we can prepared different quasi-stationary states. As an
example, we here are interested in the initial NOON state
$|\psi_0(0)\rangle=\frac{1}{\sqrt{2}}[|0,2\rangle
+(-1)^{n+1}|2,0\rangle]$ of Eqs. (9) and (6). Under the corresponding initial
conditions $b'_{0}(0)=-b'_{2}(0)=1/\sqrt{2}, b'_{1}(0)=0$, starting
form Eqs. (11-13), time evolutions of $P_{j}(t)$ are shown in Fig.
3(c) for $\omega=50, \varepsilon_{0}/\omega=2$, and $U=2$, where the
values of $P_{0}=|A_0|^{2}$ and $P_{2}=|C_0|^{2}$ are always $1/2$
and the value of $P_{1}=|B_0|^{2}$ is always $0$. Thus the
non-Floquet state of Eq. (10) actually equates the first Floquet
state of Eq. (9), $|\psi(t)\rangle=|\psi_0(t)\rangle$, due to
$c_0=1,\ c_{1}=c_{2}=0$. Such a time-dependent state possesses the
same constant population with a NOON state and is called the
quasi-NOON state thereby. For the same parameters and the initial conditions $a_{0}(0)=1/\sqrt{2},\
a_{1}(0)=0,\ a_{2}(0)=-1/\sqrt{2}$, numerical results based on
Eq. (3) are shown by
the solid lines of Fig. 3(c), which coincide with the obtained
analytical results. The results consistently demonstrate that the
entangled quasi-NOON state can be prepared by setting the initial
NOON state and applying the high-frequency driving.

It is worth noting that Eqs. (11-13) can fit arbitrary initial conditions of the system. Therefore, based on the different CDTs we can start from any initial state to construct the corresponding stationary-like state. As the considered example, for the initial conditions $b'_{0}(0)=1,b'_{1}(0)=b'_{2}(0)=0$, Eqs. (11-13) are reduced to Eqs. (14-16) which are associated with the initial paired state $|\psi(0)\rangle=|0,2\rangle$ and the final stationary-like state $|\psi(t)\rangle=e^{\frac{i\varepsilon_{0}}{\omega}\sin(\omega t)-i(n\omega
+u)t}|0,2\rangle$ of Eq. (10). If the initial state is prepared as $|\psi(0)\rangle=|2,0 \rangle$ or $|\psi(0)\rangle=(|0,2\rangle-|2,0 \rangle)/\sqrt{2}$, of course, from Eqs. (11-13) the final stationary-like state reads $|\psi(t)\rangle=e^{\frac{-i\varepsilon_{0}}{\omega}\sin(\omega t)-i(n\omega
+u)t}|2,0 \rangle$ or the quasi-NOON state $|\psi(t)\rangle=[e^{\frac{i\varepsilon_{0}}{\omega}\sin(\omega t)-i(n\omega
+u)t}|0,2 \rangle-e^{-\frac{i\varepsilon_{0}}{\omega}\sin(\omega t)-i(n\omega
+u)t}|2,0 \rangle]/\sqrt{2}$. These results will be used to design the quantum tunneling switch in next subsection.

\subsection{Quantum tunneling switch}

From the above two subsections we know that driving field affect tunneling
dynamics dramatically, so we can design the quantum tunneling switch \cite{Lu2} through the high-frequency driving field
that controls tunneling to open or close, as shown in Fig. 4. Let the two weakly interacting bosons occupy the paired state $|0,2\rangle$ initially and fix the parameters $\omega=50$ and
$U=2$. We give the scheme of quantum tunneling switch between different stationary-like states by adjusting the driving strength, namely using CDT or CCT to close or open the tunneling.

\emph{Switch of paired bosons tunneling}. In Fig. 4(a), one can see that in time interval $0\le t< t_{1}$, we let driving strength keep the value at the avoided level-crossing point $\varepsilon_{0}=2.405 \omega$, so the initially occupied state is kept, due to the CDT of the second kind in Fig. 3. At an arbitrarily given time $t=t_1$, we change the driving strength to $\varepsilon_{0}=2 \omega$ and keep this value until $t=t_3$ at which the system transits completely from state $|0,2\rangle$ to state $|2,0\rangle$, because of the CCT in this time interval. Then we return the driving strength to the initial value such that the CDT makes the final state the stationary-like state $e^{-\frac{i\varepsilon_{0}}{\omega}\sin(\omega
t)-i(n\omega+u) t}|2,0\rangle$. Thus we transport the two bosons from well 2 to well 1, through the quantum tunneling switch. The switching time is just the tunneling time $t_3-t_1\approx 125(\omega_0^{-1})=125/200$s$=0.625$s which is shown in the inset of Fig. 2(b). Note that in the considered case, Fig. 2(b) points out $P_1(t)\approx 0$.

\emph{Switch from the initial state $|2,0\rangle$ to the final quasi-NOON state}. Before $t<t_2$, we perform the similar operations with that of Fig. 4(a). At $t=t_2$ we return the driving strength to the initial value for CDT of the second kind that results in $P_0(t)\approx P_2(t)\approx 1/2$ for $t\ge t_2$, as shown in Fig. 4(b). Here we have used $t_2$ as new initial time and values of $P_j(t_2)$ for $j=0,1,2$ as the initial conditions of Eqs. (11-13) to plot $P_j(t)$ for $t>t_2$. The normalization implies $P_1(t_2)\approx 0$, because of $P_0(t_2)+ P_2(t_2)\approx 1$. Therefore we have achieved the transition from the initial state $|2,0\rangle$ to the final quasi-NOON state $|\psi(t)\rangle=e^{\frac{i\varepsilon_{0}}{\omega}\sin(\omega t)-i(n\omega +u)t}[|0,2 \rangle-e^{-2\frac{i\varepsilon_{0}}{\omega}\sin(\omega t)}|2,0 \rangle]/\sqrt{2}$. The corresponding switching time is the half tunneling time $t_2-t_1\approx 62.5(\omega_0^{-1})=0.3125$s.

In addition, making use of CDT, we can arrive at different final states with the constant probabilities $P_0(t)=P_0(\tau),\ P_2(t)=P_2(\tau)$ for $t\ge \tau$, by returning the driving strength to the avoided level-crossing point at a different fixed time $\tau \in (t_1,\ t_3)$. Clearly, if we change the interaction strength from $U=2$ to $U=\omega=50$ and adjust the driving strength either $\varepsilon_{0}=3.832 \omega$ or $\varepsilon_{0}=2 \omega$ at appropriate times, the switching time in Fig. 4(a) can be shortened to about $6(\omega_0^{-1})=0.03$s by CCT from the resonant interaction shown in Fig. 2(a).

\begin{figure}[htbp]
\centering
\includegraphics[height=1.2in,width=1.6in]{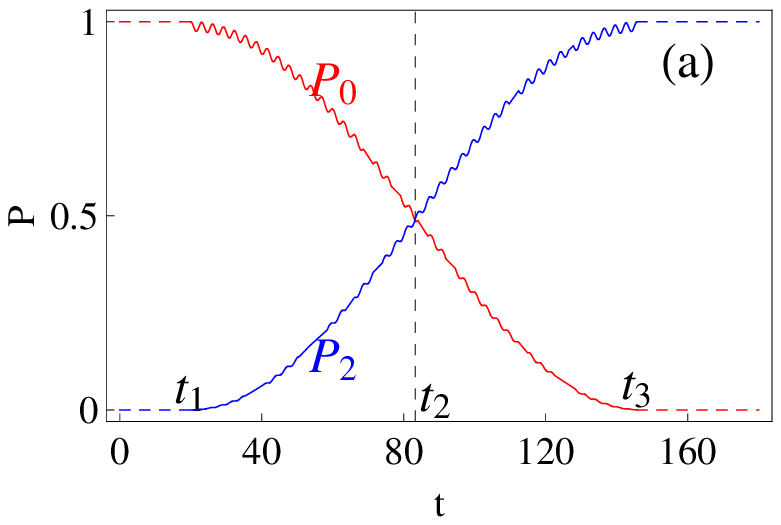}
\includegraphics[height=1.2in,width=1.6in]{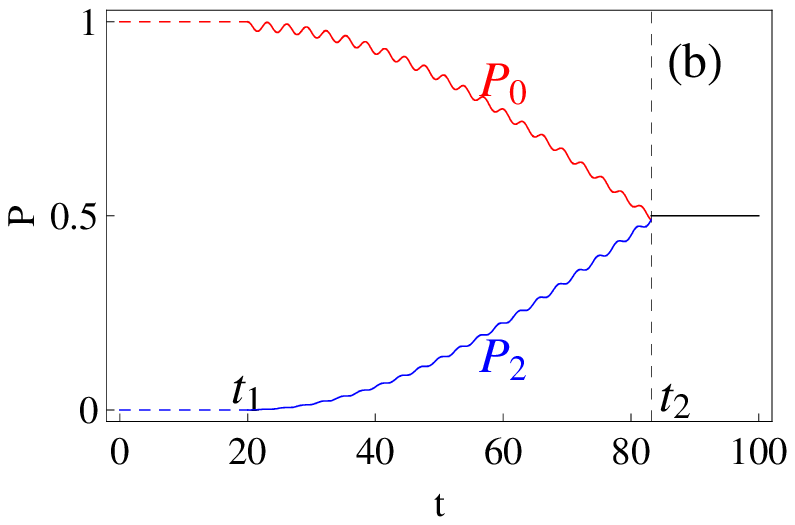}
\caption{\scriptsize{(Color online) Plots showing the schemes of
quantum tunneling switch, through the time evolutions of
probabilities $P_{0}(t)$ and $P_{2}(t)$ from Eqs. (11-16) for
$\omega=50, \gamma=0.5$. (a) In the time intervals $0\le t< t_{1}$
and $t\ge t_3$, the parameter values are taken at the avoided
level-crossing point $(\varepsilon_{0}/\omega,\ U)=(2.405,\ 2)$, and
in the time interval $t_1\leq t< t_3$, only the driving strength is
changed to $\varepsilon_{0}=2 \omega$. (b) In the time intervals
$0\le t< t_{1}$ and $t\ge t_2$, the parameter values are taken at
the same avoided level-crossing point with that of Fig. 4(a), and
the driving strength is kept as $\varepsilon_{0}=2 \omega$ in
$t_1\leq t< t_2$. Here $t_{1}=20,\ t_{2}=82.5$ and $t_3=145 $ are
associated with the approximate tunneling time
$t_3-t_1=125(\omega_0^{-1})$s and its half
$t_2-t_1=125/2=62.5(\omega_0^{-1})$.}}
\end{figure}

\section{Conclusion}

We have considered two repulsive bosons confined in a periodically
driven double well and investigated the interplay between
interparticle interaction and external field and its application to
coherent control of quantum tunneling dynamics. Under high-frequency
limit, we obtained Floquet quasienergies, Floquet states of
invariant population. A full solution of the Schr\"odinger equation is constructed as a general non-Floquet state of slowly variant
population, which is coherent superposition of the three Floquet
states. The analytical quasienergy spectra exhibit the
level-crossing for resonant interaction, the avoided crossings of
partial levels for non-resonant interaction, and the translations of
level-crossing points caused by the photon resonance.
Exploiting the coherent non-Floquet states, we obtain time
evolutions of the bosonic population, where beyond the
level-crossing points, resonant interaction results in CCT which
enhances tunneling rate of the paired bosons. For weak interaction,
two repelling bosons tunnel between wells as pair and fall on a NOON
state periodically. Three different kinds of CDT are found for the
three-level crossing, avoided crossing of partial levels and arbitrary values of the system parameters, respectively. At the level-crossing points,
tunneling of two bosons between wells is suppressed completely that
means occurrence of the first kind CDT. Such a CDT needs larger
driving strength for a stronger interaction. The avoided crossing of
partial levels means the two-level crossing, so the superposition
state of two Floquet states with the same energy describes CDT of
the second kind. If the system is prepared in a single Floquet state
initially, CDT of the third kind occurs for arbitrary values of the system parameters in high-frequency regime, including those at the level-crossing and -uncrossing points. The CDTs of different kinds lead to different stationary-like states in which the probability amplitudes
are time-dependent and the corresponding probabilities are
time-independent.

In order to confirm the analytical results, we make direct numerical simulations and demonstrate perfect agreement between both results.
As an application of these results, we present an interesting design scheme of quantum tunneling switch, through the paired-particle tunnelings between different stationary-like states which contain the entangled quasi-NOON state, by using CDT or CCT to close or open the quantum tunneling. Such a scheme could be useful for controlling quantum tunneling of two bosons in a double well.

\section{Acknowledgements}
This work was supported by the NNSF of China under Grant No.
11175064, the Construct Program of the National Key Discipline, the
PCSIRTU of China (IRT0964), and the Hunan Provincial NSF (11JJ7001).


\begin{thebibliography}{999}

\bibitem{Milena} M. Grifoni and P. H\"{a}nggi, Phys. Rep. 304, 229
(1998).
\bibitem{Kohler} S. Kohler, J. Lehmann, and P. H\"{a}nggi, Phys. Rep. 406,379
(2005).
\bibitem{P} P. Kr\'al, Rev. Mod. Phys. 79, 53 (2007).
\bibitem{E.Kierig} E. Kierig, U. Schnorrberger, A. Schietinger, J. Tomkovic,
and M. k. Oberthaler, Phys. Rev. Lett 100, 190405 (2008).
\bibitem{G.Della} G. D. Valle, M. Ornigotti, E. Cianci, V. Foglietti, P. Laporta, and S. Longhi, Phys. Rev. Lett 98, 263601 (2007).

\bibitem{F.Grossmann} F. Grossmann, T. Dittrich, P. Jung, and P. H\"{a}nggi, Phys. Rev. Lett 67. 516 (1991).
\bibitem{Xie} Q. Xie and W. Hai,  Phys. Rev. A75, 015603 (2007).

\bibitem{Gong} J. Gong, L. Morales-Molina, and P. H\"anggi, Phys. Rev. Lett.
103, 133002 (2009).
\bibitem{Longhi1} S. Longhi, Phys. Rev. A 86, 044102 (2012).
\bibitem{Hai1} W. Hai, K. Hai and Q. Chen, Phys. Rev. A 87, 023403 (2013).

\bibitem{Lin} W. A. Lin and L. E. Ballentine, Phys. Rev. Lett. 65, 2927(1990).
\bibitem{LuG}  G. Lu, W. Hai, H. Zhong, and Q. Xie, Phys. Rev. A81, 063423 (2010);  G. Lu, W. Hai and H. Zhong, Phys. Rev. A80, 013411 (2009).
\bibitem{C.Sias} C. Sias, H. Lignier, Y. P. Singh, A. Zenesini, D. Ciampini,O. Morsch, and E. Arimondo, Phys. Rev. Lett 100, 040404 (2008).
\bibitem{Q} Q. Xie, S. Rong, H. Zhong, G. Lu, and W. Hai, Phys. Rev. A 82, 023616 (2010).

\bibitem{C.E} C. E. Creffield and T. S. Monteiro, Phys. Rev. lett 96, 210403 (2006).
\bibitem{Longhi2} S. Longhi and G. D. Valle,  Phys. Rev. A 86, 042104 (2012).


\bibitem{Creffield} C. E. Creffield, Phys. Rev. Lett. 99, 110501 (2007).
\bibitem{Romero} O. Romero-Isart and J. J. Garci\'a-Ripoll, Phys.Rev.A76, 052304 (2007).
\bibitem{Hai} K. Hai, W. Hai and Q. Chen, Phys. Rev. A82, 053412 (2010).
\bibitem{M.Esmann} M. Esmann, N. Teichmann, and C. Weiss, Phys. Rev. A 83, 063634 (2011)

\bibitem{D.S} D. S. Murphy, J . F. McCann, J. Goold and Th. Busch, Phys. Rev. A 76, 053616 (2007).
\bibitem{P.Cheinet} P. Cheinet, S. Trotzky, M. Feld, U. Schnorrberger, M. Moreno-Cardoner, S. Folling, and I. Bloch, Phys. Rev. Lett 101, 090404 (2008).
\bibitem{Budhaditya}  B. Chatterjee, I. Brouzos, S. Z\"{o}llner, and P. Schmelcher, Phys. Rev. A 82, 043619 (2010).

\bibitem{Sascha} S. Z\"{o}llner, H. D Meyer, and P. Schmelcher, Phys. Rev. Lett 100, 040401 (2008); Phys. Rev. A 78, 013621 (2008).

\bibitem{Longhi} S. Longhi, Phys. Rev. A 83, 043835 (2011).

\bibitem{Sascha2} S. F\"{o}lling, S. Trotzky, P. Cheinet, M. Feld, R. Saers, A. Widera, T. M\"{u}ller, and I. Bloch, Nature (London) 448, 1029 (2007).
\bibitem{M.Z} M. Zwierlein, J. Abo-Shaeer, A. Schirotzek, C. Schunck, and
W. Ketterle, Nature (London) 435, 1047 (2005).

\bibitem{Chen} Y. A. Chen, X. H. Bao, Z. S. Yuan, S. Chen, B. Zhao, and J. W.
Pan, Phys. Rev. Lett. 104, 043601 (2010).
\bibitem{K.Stiebler} K. Stiebler, B. Gertjerenten, N. Teichmann and C. Weiss, J. Phys. B 44, 055301 (2011).

\bibitem{Yosuke} Y. Kayanuma and K. Saito, Phys. Rev. A 77, 010101(R) (2008).

\bibitem{D.Jaksch} D. Jaksch, C. Bruder, and J. I. Cirac, Phys. Rev. Lett 81, 3108 (1998).
\bibitem{Jinasundera} T. Jinasundera, C. Weiss, and M. Holthaus, Chem. Phys. 322, 118 (2006).
\bibitem{MHolthaus} M. Holthaus, Phys. Rev. A 64, 011601(R) (2008).

\bibitem{Sambe} H. Sambe, Phys. Rev. A 7, 2203 (1973).
\bibitem{Lu} G. Lu, W. Hai, and Q. Xie, Phys. Rev. A 83, 013407 (2011).

\bibitem{Lu2} G. Lu and W. Hai, Phys. Rev. A 83, 053424 (2011).

\end{thebibliography}
\end{document}